\documentclass[aps,nofootinbib,commonaddress,preprint,showpacs]{revtex4}
\usepackage{amsmath}
\usepackage{amssymb}
\usepackage{graphicx}
\usepackage{color}

\begin{document}

\title{A new type of surface waves in a fully degenerate quantum plasma}
\author{Yu.~Tyshetskiy}
\email{y.tyshetskiy@physics.usyd.edu.au}
\author{S.V.~Vladimirov}
\affiliation{School of Physics, The University of Sydney, NSW 2006, Australia}
\affiliation{Metamaterials Laboratory, National Research University of Information Technology, Mechanics, and Optics, St Petersburg 199034, Russia}
\author{R.~Kompaneets}
\affiliation{School of Physics, The University of Sydney, NSW 2006, Australia}

\date{\today}
\received{}

\begin{abstract}
We study the response of a semi-bounded one-component fully degenerate electron plasma to an initial perturbation in the electrostatic limit. We show that the part of the electric potential corresponding to surface waves in such plasma can be represented, at large times, as the sum of two terms, one term corresponding to ``conventional'' (Langmuir) surface waves and the other term representing a new type of surface waves resulting from specific analytic properties of degenerate plasma's dielectric response function. These two terms are characterized by different oscillation frequencies (for a given wave number), and, while the ``conventional'' term's amplitude decays exponentially with time, the new term is characterized by a slower, power-law decay of the oscillation amplitude and is therefore dominant at large times.
\end{abstract}

						%52.35.-g: Waves, oscillations, and instabilities in plasmas and intense beams
						%52.65.-y: Plasma simulation
						%52.25.Dg: Plasma kinetic equations 
						%52.25.Mq	Dielectric properties (of plasmas)
						%05.30.-d: Quantum statistical mechanics
						%41.20.Cv  Electrostatics; Poisson and Laplace equations, boundary-value problems
						
\maketitle

\section{Introduction}
Surface plasma waves (SPW) are collective oscillations supported by bounded media, which propagate along an interface of two media with different signs of the real part of dielectric response function. What distinguishes them from volume plasma waves (which can propagate in both unbounded and bounded media) is that their field is localized near the interface along which they propagate. They also have different spectral and attenuation properties than volume plasma waves~\cite{ABR_book}.

Surface plasma waves have been studied extensively since their theoretical prediction~\cite{Ritchie_1957} and experimental detection~\cite{Powell_Swan_1959_1,Powell_Swan_1959_2,Otto_1968,Kretschmann_1968} in 1950s and 1960s. There has been a significant advance in theoretical and experimental investigations of surface plasma waves in various bounded plasma structures, both in the field of plasma science [see Refs~\cite{Vladimirov_progress_1994} and references therein] and in the fields of condensed matter and surface science [see, e.g., a review~\cite{Pitarke_etal_2007}]. Currently, there is a renewed interest in surface plasma waves due to their ability to concentrate light in subwavelength structures, enabling to create SPW-based circuits that can couple photonics and electronics at nanoscale. This offers a route to faster and smaller devices, and opens up possibilities to new technologies employing surface plasma waves~\cite{Plasmonics_Science_2010,Zheludev_etal_2008,Noginov_etal_2009,Lasers_go_nano}.

These developments require a solid understanding of SPW properties in bounded metallic and semiconductor structures.
The properties of surface plasma waves are defined, among other things, by the dielectric properties of the medium that sustains them. The latter are often (e.g., in metals, for which the electrons are strongly degenerate) significantly affected by the quantum nature of the charge carriers~\cite{Fortov_book}. This can affect the properties of SPW in a non-trivial way, via modification of analytic properties of the medium response. In particular, quantum effects (due to Pauli blocking and overlapping wave functions of free charge carriers in the medium~\cite{Eliasson_Shukla_RevModPhys_2011}), when significant, can modify the dispersion, damping~\cite{Tyshetskiy_etal_PoP_2012} and spatial attenuation of SPW~\cite{Vladimirov_Kohn_1994} supported by a bounded medium. 

Recently, the properties of surface plasma waves in a semi-bounded degenerate plasma have been analyzed using quantum hydrodynamical approach~\cite{Lazar_etal_2007}, and a soon after -- with a more rigorous kinetic approach~\cite{Tyshetskiy_etal_PoP_2012}. In particular, the effects of quantum recoil and quantum degeneracy of plasma electrons on SPW properties have beeen analyzed. 

In this paper, we show another important consequence of quantum degeneracy of electrons on SPW properties, exemplified by a simple case of SPW in a semi-bounded collisionless plasma with degenerate electrons. Namely, we show that such system supports \textit{two} types of SPW, with different frequencies and qualitatively different temporal attenuation, in contrast to a case of non-degenerate semi-bounded plasma that only supports one (``conventional'') type of SPW~\cite{Guernsey_1969}. The new type of surface oscillations predicted here is shown to become dominant over the ``conventional'' surface oscillations at large times, and should therefore become observable, e.g., by analyzing a spectrum of the reflected light in experimental setups for excitation of surface plasmons in thin metal films by an incident light using Otto or Kretschmann configurations~\cite{Otto_1968,Kretschmann_1968}. At large wavelengths, the frequency difference between these two types of surface oscillations approaches a third of the metal's plasma frequency, and thus the absorption lines in the reflected light spectrum, corresponding to excitation of surface waves of these two types, should be clearly separated and detectable.

\section{Method}
\subsection{Model and assumptions}
We consider a semi-bounded, nonrelativistic collisionless plasma with degenerate mobile electrons ($T_e\ll\epsilon_F$, where $T_e$ is the electron temperature in energy units, $\epsilon_F=\hbar^2(3\pi^2n_e)^{2/3}/2 m_e$ is the electron Fermi energy), and immobile ions; the equilibrium number densities of electrons and ions are equal, $n_{0e}=n_{0i}=n_0$ (quasineutrality). The plasma is assumed to be confined to a region $x<0$, with mirror reflection of plasma particles at the boundary $x=0$ separating the plasma from a vacuum at $x>0$. 

We will look at SPWs in the non-retarded limit, when their phase velocity is small compared with the speed of light. In this limit, the SPW field is purely electrostatic, hence we can restrain ourselves to considering only electrostatic oscillations in the considered system. Following the discussion of Ref.~\cite{Tyshetskiy_etal_PoP_2012}, we adopt here the quasiclassical kinetic description of plasma electrons in terms of the 1-particle distribution function $f(\mathbf{r,v},t)=f(x,\mathbf{r}_\parallel,v_x,\mathbf{v}_\parallel,t)$~\cite{Vlad_Tysh_UFN_2011} (where $\mathbf{r}_\parallel$ and $\mathbf{v}_\parallel$ are, respectively, the components of $\mathbf{r}$ and $\mathbf{v}$ parallel to the boundary, and $x$ and $v_x$ are the components of $\mathbf{r}$ and $\mathbf{v}$ perpendicular to the boundary), whose evolution is described by the kinetic equation
\begin{equation}
\frac{\partial f}{\partial t} + v_x\frac{\partial f}{\partial x} + \mathbf{v}_\parallel\cdot\frac{\partial f}{\partial\mathbf{r}_\parallel} - \frac{e}{m_e}\left(\frac{\partial\phi}{\partial x}\frac{\partial f}{\partial v_x} + \frac{\partial\phi}{\partial\mathbf{r}_\parallel}\cdot\frac{\partial f}{\partial\mathbf{v}_\parallel}\right) = 0, \label{eq:Vlasov}
\end{equation}
where the electrostatic potential $\phi(x,\mathbf{r}_\parallel,t)$ is defined by the Poisson's equation
\begin{equation}
-\nabla^2\phi = 4\pi e\left[\int{f(\mathbf{r,v},t)d^3\mathbf{v}} - n_0\right]. \label{eq:Poisson}
\end{equation}

% In the absence of fields, the equilibrium distribution function of plasma electrons $f_0(\mathbf{v})$ is defined by the Pauli's exclusion principle for fermions, and for low electron temperatures $T_e/\epsilon_F\ll 1$ (where $T_e$ is the electron temperature, $\epsilon_F=m_ev_F^2/2=(3\pi^2\hbar^3 n_e)^{2/3}/2m_e$ is the electron Fermi energy) it becomes
% \begin{equation}
% f_0(v)=\begin{cases}
% \frac{3 n_0}{4\pi v_F^3} = \frac{2 m_e^3}{(2\pi\hbar)^3} &\mbox{if }v\leq v_F, \\
% 0 &\mbox{if }v > v_F,
% \end{cases}  \label{eq:f0}
% \end{equation}
% corresponding to fully degenerate electron distribution.

In the absence of fields, the equilibrium distribution function of plasma electrons $f_{0}(\mathbf{v})$ is defined by an isotropic Fermi-Dirac distribution, which in the limit $T_e\ll\epsilon_F$ reduces to
\begin{equation}
f_{0}(v)=\frac{2 m_e^3}{(2\pi\hbar)^3}\left\{1 + \exp\left[\frac{m_e v^2/2 - \epsilon_F(n_e)}{T_e}\right] \right\}^{-1} = \frac{2 m_e^3}{(2\pi\hbar)^3} \sigma\left[v_F(n_e)-v\right],  \label{eq:f0}
\end{equation}
where $v_F(n_e)=\sqrt{2\epsilon_F(n_e)/m_e}$ is the electron Fermi velocity, $\sigma(x)$ is the Heaviside step function.

The condition of mirror reflection of plasma electrons off the boundary at $x=0$ implies
\begin{equation}
f(x=0,\mathbf{r}_\parallel,-v_x,\mathbf{v}_\parallel,t) = f(x=0,\mathbf{r}_\parallel,v_x,\mathbf{v}_\parallel,t).  \label{eq:boundary}
\end{equation}

\subsection{Initial value problem}
% \textcolor{blue}{\textit{Formulate the initial value problem: initial perturbation and its evolution in dimensionless variables (from Guernsey). Discuss the terms in the solution.}}

We now introduce a small initial perturbation $f_p(x,\mathbf{r}_\parallel,v_x,\mathbf{v}_\parallel,t=0)$ to the equilibrium electron distribution function $f_0(v)$, $|f_p(x,\mathbf{r}_\parallel,v_x,\mathbf{v}_\parallel,t=0)|\ll f_0(v)$, and use the kinetic equation (\ref{eq:Vlasov}) to study the resulting evolution of the system's charge density $\rho(x,\mathbf{r}_\parallel,t)=e\left[\int{f(x,\mathbf{r_\parallel,v},t)d^3\mathbf{v}}-n_0\right]$, and hence of the electrostatic potential $\phi(x,\mathbf{r}_\parallel,t)$ defined by (\ref{eq:Poisson}). Introducing the dimensionless variables $\Omega=\omega/\omega_p$, $\mathbf{K}=\mathbf{k}\lambda_F$, $\mathbf{V}=\mathbf{v}/v_F$, $X=x/\lambda_F$, $\mathbf{R}_\parallel=\mathbf{r}_\parallel/\lambda_F$, $\lambda_F=v_F/\sqrt{3}\omega_p$, $\omega_p=(4\pi e^2 n_0/m_e)^{1/2}$, and following Guernsey~\cite{Guernsey_1969}, the solution of the formulated initial value problem for $\rho(X,\mathbf{R}_\parallel,T)$ with the boundary condition (\ref{eq:boundary}) is
\begin{eqnarray}
\rho(X,\mathbf{R}_\parallel,T) &=& en_0\tilde{\rho}(X,\mathbf{R}_\parallel,T), 
\end{eqnarray}
where
\begin{eqnarray}
\tilde{\rho}(X,\mathbf{R}_\parallel,T) &=& \frac{1}{(2\pi)^3}\int_{-\infty}^{+\infty}{dK_x{\ e}^{i K_x X}\int{d^2\mathbf{K}_\parallel}{\ e}^{i\mathbf{K}_\parallel\cdot\mathbf{R}_\parallel}\ \tilde{\rho}_{\mathbf{k}}(T)}, \\
\tilde{\rho}_{\mathbf{K}}(T) &=& \frac{1}{2\pi}\int_{i\sigma-\infty}^{i\sigma+\infty}{\tilde{\rho}(\Omega,\mathbf{K}){\ e}^{-i\Omega T}d\Omega},\ \rm{with\ }\sigma>0. \label{eq:inv_Laplace} 
\end{eqnarray}
The integration in (\ref{eq:inv_Laplace}) is performed in complex $\Omega$ plane along the horizontal contour that lies in the upper half-plane ${\rm Im}(\Omega)=\sigma>0$ above all singularities of the function $\tilde{\rho}(\Omega,\mathbf{K})$. The function $\tilde{\rho}(\Omega,\mathbf{K})$, defined as the Laplace transform of $\tilde\rho_\mathbf{K}(T)$:
\begin{equation}
\tilde{\rho}(\Omega,\mathbf{K}) = \int_0^\infty{\tilde\rho_\mathbf{K}(T)\ e^{i\Omega T} dT}, \label{eq:Laplace}
\end{equation}
is found to be
\begin{eqnarray} 
\tilde{\rho}(\Omega,\mathbf{K}) &=& \frac{i}{\varepsilon(\Omega,K)}\int{d^3\mathbf{V}\frac{G(\mathbf{V,K})}{\Omega-\sqrt{3} \mathbf{K\cdot V}}} \nonumber \\
&+&\frac{iK_\parallel}{2\pi\zeta(\Omega,K_\parallel)}\left[1-\frac{1}{\varepsilon(\Omega,K)}\right]\int_{-\infty}^{+\infty}{\frac{dK_x'}{{K^\prime}^2\ \varepsilon(\Omega,K')}\int{d^3\mathbf{V}\frac{G(\mathbf{V},\mathbf{K^\prime})}{\Omega-\sqrt{3}\mathbf{K'\cdot V}}}},  \label{eq:rho(w,k)} 
\end{eqnarray}
where the Fourier transforms $G(\mathbf{V,K})$ and $G(\mathbf{V,K'})$ of the (dimensionless) initial perturbation are defined by
\begin{eqnarray}
G(\mathbf{V,K}) &=& \int_{-\infty}^{+\infty}{dX {\ e}^{-i K_x X}\  \tilde{g}(X,V_x,\mathbf{V}_\parallel,\mathbf{K}_\parallel)}, 
% \\
% G(\mathbf{V,K'}) &=& \int_{-\infty}^{+\infty}{dX {\ e}^{-i K_x' X}\  \tilde{g}(X,V_x,\mathbf{V}_\parallel,\mathbf{K}_\parallel)}, 
\end{eqnarray}
with
\begin{eqnarray}
\tilde{g}(X,V_x,\mathbf{V}_\parallel,\mathbf{K}_\parallel) &=& \int{d^2\mathbf{R}_\parallel {\ e}^{-i\mathbf{K_\parallel\cdot R_\parallel}}\  \tilde{f_p}(X,\mathbf{R}_\parallel,V_x,\mathbf{V}_\parallel,0)}, \\
\tilde{f_p}(X,\mathbf{R}_\parallel,V_x,\mathbf{V}_\parallel,0) &=& \frac{v_F^3}{n_0} f_p(X,\mathbf{R}_\parallel,V_x,\mathbf{V}_\parallel,0),
\end{eqnarray}
where $K_\parallel=|\mathbf{K}_\parallel|$, $K=|\mathbf{K}|$, $\mathbf{K}=(K_x,\mathbf{K}_\parallel)$, $K'=|\mathbf{K'}|$, and $\mathbf{K'}=(K_x',\mathbf{K}_\parallel)$.
The functions $\varepsilon(\Omega,K)$ and $\zeta(\Omega,K_\parallel)$ in (\ref{eq:rho(w,k)}) are defined (for ${\rm Im}(\Omega)>0$) as follows:
\begin{eqnarray}
\varepsilon(\Omega,K) &=& 1 - \frac{1}{\sqrt{3}K^2}\int{\frac{\mathbf{K}\cdot\partial\tilde{f}_0(\mathbf{V})/\partial\mathbf{V}}{\Omega-\sqrt{3}\mathbf{K\cdot V}}d^3\mathbf{V}}, \label{eq:epsilon} \\
\zeta(\Omega,K_\parallel) &=& \frac{1}{2} + \frac{K_\parallel}{2\pi}\int_{-\infty}^{+\infty}{\frac{dK_x}{K^2\ \varepsilon(\Omega,K)}}, \label{eq:zeta}
\end{eqnarray} 
with 
\[
\tilde{f}_0(\mathbf{V}) = \frac{v_F^3}{n_0}f_0(\mathbf{V}) = \frac{v_F^3}{n_0}\left.f_0(\mathbf{v})\right|_{\mathbf{v}=v_F\mathbf{V}}. 
\]
For fully degenerate plasma with electron distribution (\ref{eq:f0}), the function $\varepsilon(\Omega,K)$ becomes (for ${\rm Im}(\Omega)>0$)~\cite{Gol'dman_1947,ABR_book}:
\begin{equation}
\varepsilon(\Omega,K) = 1 + \frac{1}{K^2}\left[1-\frac{\Omega}{2\sqrt{3}K}\ln\left(\frac{\Omega+\sqrt{3}K}{\Omega-\sqrt{3}K}\right)\right],\ \ {\rm Im}(\Omega)>0,  \label{eq:epsilon_degenerate}
\end{equation}
where $\ln(z)$ is the principal branch of the complex natural logarithm function.

Note that the solution~(\ref{eq:rho(w,k)}) differs from the corresponding solution of the transformed Vlasov-Poisson system for infinite (unbounded) uniform plasma only in the second term involving $\zeta(\Omega,K_\parallel)$; indeed, this term appears due to the existance of boundary at $x=0$.

% \textcolor{blue}{\textit{Say that the poles $\varepsilon(\Omega,K)=0$ of $\tilde{\rho}(\Omega,\mathbf{K})$ in (\ref{eq:inv_Laplace}) define the properties of volume waves, and the poles $\zeta(\Omega,K_\parallel)=0$ of $\tilde{\rho}(\Omega,\mathbf{K})$ in (\ref{eq:inv_Laplace}) define the properties of surface waves, as shown by Guernsey. Below we discuss the analytic properties and evaluation of $\zeta(\Omega,K_\parallel)=0$, from which the dispersion and damping of the surface wave solution will be obtained.}}

The definition (\ref{eq:Laplace}) of the function $\tilde\rho(\Omega,\mathbf{K})$ of complex $\Omega$ has a sense (i.e., the integral in (\ref{eq:Laplace}) converges) only for ${\rm Im}(\Omega)>0$. Yet the long-time evolution of $\tilde{\rho}_\mathbf{k}(T)$ is obtained from (\ref{eq:inv_Laplace}) by displacing the contour of integration in complex $\Omega$ plane from the upper half-plane ${\rm Im}(\Omega)>0$ into the lower half-plane ${\rm Im}(\Omega)\leq0$~\cite{Landau_1946}. This requires the definition of $\tilde\rho(\Omega,\mathbf{K})$ to be extended to the lower half-plane, ${\rm Im}(\Omega)\leq 0$, by analytic continuation of (\ref{eq:rho(w,k)}) from ${\rm Im}(\Omega)>0$ to ${\rm Im}(\Omega)\leq 0$. Hence, the functions
\begin{equation}
I(\Omega,\mathbf{K})\equiv\int{d^3\mathbf{V}\frac{G(\mathbf{V,K})}{\Omega-\sqrt{3}\mathbf{K\cdot V}}},  \label{eq:I}
\end{equation}
$\varepsilon(\Omega,K)$, and $\zeta(\Omega,K_\parallel)$ that make up the function $\tilde\rho(\Omega,\mathbf{K})$, must also be analytically continued into the lower half-plane of complex $\Omega$, thus extending their definition to the whole complex $\Omega$ plane. With thus continued functions, the contributions to the inverse Laplace transform (\ref{eq:inv_Laplace}) at large times $T\gg 1$ are of three sources~\cite{Guernsey_1969}: 
\begin{enumerate}
\item Contributions from the singularities of $I(\Omega,\mathbf{K})$ in the lower half of complex $\Omega$ plane (defined solely by the initial perturbation $G(\mathbf{V,K})$); with some simplifying assumptions about the initial perturbation~\cite{Guernsey_1969} these contributions are damped in a few plasma periods and can be ignored. 
\item Contribution of singularities of $1/\varepsilon(\Omega,K)$ in the lower half of complex $\Omega$ plane, of two types: (i) residues at the poles of $1/\varepsilon(\Omega,K)$, which give the volume plasma oscillations~\cite{Guernsey_1969}, and (ii) integrals along branch cuts (if any) of $1/\varepsilon(\Omega,K)$ in the lower half-plane of complex $\Omega$, which can lead to non-exponential attenuation of the volume plasma oscillations~\cite{Hudson_1962,Krivitskii_Vladimirov_1991}.
\item Contribution into (\ref{eq:inv_Laplace}) of singularities of $1/\zeta(\Omega,K_\parallel)$ in the lower half of complex $\Omega$ plane, of two types: (i) residues at the poles of $1/\zeta(\Omega,K_\parallel)$, corresponding to the surface wave solutions of the initial value problem in the considered system~\cite{Guernsey_1969,Tyshetskiy_etal_PoP_2012}, and (ii) integrals along branch cuts (if any) of $1/\zeta(\Omega,K_\parallel)$ in the lower half-plane of complex $\Omega$. 
\end{enumerate}

\begin{figure}
\centerline{\includegraphics[width=2.7in]{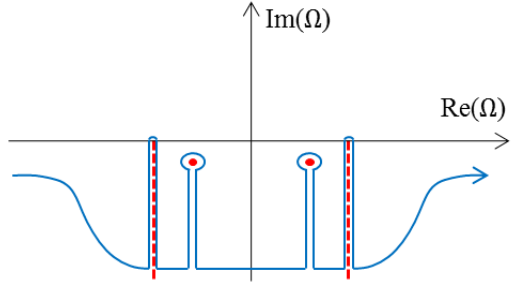}}% Images in 100% size
\caption{A sketch of the deformed integration path in (\ref{eq:inv_Laplace}) in complex $\Omega$ plane, with contributions of poles (solid circles) and branch cuts (dashed lines) of $1/\zeta(\Omega,K_\parallel)$. The singularities due to $1/\varepsilon(\Omega,K)$ are not shown, but they also contribute to (\ref{eq:inv_Laplace}), yielding volume plasma waves.}
\label{fig:sketch_integration_omega}
\end{figure}

Below we consider the latter contributions from poles and branch cuts of $1/\zeta(\Omega,K_\parallel)$ in the lower half-plane of complex $\Omega$, as illustrated in Fig.~\ref{fig:sketch_integration_omega}, and show that they yield two types of electrostatic surface oscillations with different frequencies and qualitatively different temporal attenuation. Note that the contribution of the horizontal part of the integration contour decays faster than the above contributions of singularities, and is thus negligible at large times $T\gg 1$.

\section{Two types of surface oscillations} 
\subsection{Contribution of poles of $1/\zeta(\Omega,K_\parallel)$}
The contribution of poles of $1/\zeta(\Omega,K_\parallel)$ into (\ref{eq:inv_Laplace}) leads to exponentially damped surface oscillations~\cite{Tyshetskiy_etal_PoP_2012}
\begin{equation}
\tilde\rho_{\mathbf{K}}^{\rm (poles)}(T)\propto e^{-|\Gamma_s|T}\cos\left(\Omega_s T\right), \label{eq:rho(T)_poles}
\end{equation}
with frequency $\Omega_s=\omega_s/\omega_p$ and damping rate $\Gamma_s=\gamma_s/\omega_p$ obtained from the dispersion equation $\zeta(\Omega,K_\parallel)=0$.
%and shown in Fig.~\ref{fig:omega_s+gamma_s}. 
The frequency asymptotes are
\begin{eqnarray}
\Omega_s &\approx& \frac{1}{\sqrt{2}}\left(1+0.95 K_\parallel\right),\ \rm{for\ }K_\parallel\ll 1  \label{eq:Omega_Kz<<1} \\
\Omega_s &\approx& \sqrt{3} K_\parallel \left(1+2\exp\left[-2-4K_\parallel^2\right]\right),\ \rm{for\ }K_\parallel\gg 1\ (\rm zero\ sound) \label{eq:dispersion_Kz>>1}
\end{eqnarray}
The absolute value of the damping rate is a nonmonotonic function of $K_\parallel$. At small $K_\parallel$, it increases linearly with $K_\parallel$,
\begin{equation}
\left|\Gamma_s(K_\parallel)\right|\approx 2.1\sqrt{3}\cdot 10^{-2} K_\parallel,  \label{eq:Gamma_Kz<<1}
\end{equation}
reaches maximum $|\Gamma_s|\approx 6.2\cdot 10^{-3}$ at $K_\parallel\approx 0.4$, and then quickly decreases at $K_\parallel>0.4$. Since the maximum growth rate is small, the surface oscillations due to the poles of $1/\zeta(\Omega,K_\parallel)$ are weakly damped at all wavelengths~\cite{Tyshetskiy_etal_PoP_2012}.

% \begin{figure}
% \centerline{\includegraphics[width=2.5in]{dispersion.eps} \includegraphics[width=2.5in]{damping.eps}}% Images in 100% size 
% \caption{(Color online) Frequency $\Omega_s(K_\parallel)$ (left panel) and the absolute value of damping rate $\left|\Gamma_s(K_\parallel)\right|$ (right panel) of surface oscillations due to the poles of $1/\zeta(\Omega,K_\parallel)$. The lines on the left panel show: the long-wave limit $1/\sqrt{2}$ (dotted line), the long-wave asymptote (\ref{eq:Omega_Kz<<1}) (dot-dashed blue line), and the short-wave asymptote (\ref{eq:dispersion_Kz>>1}) corresponding to zero sound propagating along the boundary (dashed red line). The dotted line on the right panel shows the long-wave asymptote (\ref{eq:Gamma_Kz<<1}) of the damping rate.}
% \label{fig:omega_s+gamma_s}
% \end{figure}

\subsection{Contribution of branch cuts of $1/\zeta(\Omega,K_\parallel)$ \label{sec:cuts}}
For degenerate plasma, the analytically continued function $\zeta(\Omega, K_\parallel)$ has two branching points on the real axis of the complex $\Omega$ plane at $\Omega=\pm\Omega_v(K_\parallel)$, where $\Omega_v(K_\parallel)\in\mathbb{R}$ is the solution of equation
\begin{equation}
\varepsilon(\Omega,K_\parallel)=\left.\varepsilon(\Omega,K)\right|_{K_x=0}=0,  \label{eq:Omega_v}
\end{equation}
with the corresponding branch cuts going down into the ${\rm Im}(\Omega)<0$ part of the complex $\Omega$ plane, as schematically shown in Fig.~\ref{fig:sketch_integration_omega} (see Appendix~\ref{app:branching}). Let us consider the contribution of the integration along these branch cuts into the inverse Laplace transform (\ref{eq:inv_Laplace}). The branching points lie above the poles $\Omega_s-i|\Gamma_s|$ of $1/\zeta$ (since the latter lie below the real axis of the $\Omega$ plane), therefore we can expect the contribution of the integration along the branch cuts into (\ref{eq:inv_Laplace}) to be at least as important as the contribution of the poles, if not to exceed it.

\begin{figure}
\centerline{\includegraphics[width=3.5in]{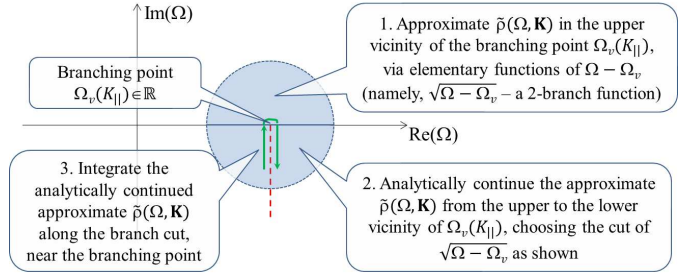}}% Images in 100% size
\caption{Evaluation scheme for the integrals along branch cuts, as outlined in Sec.~\ref{sec:cuts}.}
\label{fig:anal_cont}
\end{figure}

At large times $T\gg 1$, the main contribution into the integrals along the branch cuts comes from the small vicinity of the branching points, so it suffices to approximate the second term of (\ref{eq:rho(w,k)}) near the branching points in the lower semiplane of complex $\Omega$. This can be done in two steps:
\begin{enumerate}
\item Approximate $\tilde\rho(\Omega,\mathbf{K})$ defined by (\ref{eq:rho(w,k)}) in the upper vicinities of the branching points, in terms of elementary functions; the approximate function should have the same branching points as the original one.
\item Analytically continue these approximations into the lower vicinities of the branching points, choosing the branch cuts to go downwards from the branching points.
\end{enumerate}
Then we can perform the integration of thus obtained approximations along the branch cuts in the vicinity of the branching points. This scheme is sketched in Fig.~\ref{fig:anal_cont}

The function (\ref{eq:rho(w,k)}) in the upper vicinity of the right branching point $\Omega=+\Omega_v(K_\parallel)$ can be approximated as~(see Appendix~\ref{app:rho_approximate})
\begin{eqnarray}
\tilde\rho(\Omega,\mathbf{K})&\approx& i\frac{I(\Omega_v,\mathbf{K})}{\left.\varepsilon(\Omega,K)\right|_{\Omega\approx\Omega_v}} \nonumber \\
&+&i\left[1-\frac{1}{\left.\varepsilon(\Omega,K)\right|_{\Omega\approx\Omega_v}}\right]\frac{\left.I(\Omega_v,\mathbf{K})\right|_{K_x=0}}{1+\sqrt{\alpha(K_\parallel)\beta(K_\parallel)}K_\parallel\sqrt{\Omega-\Omega_v}}, \label{eq:rho_approx}
\end{eqnarray}
where $I(\Omega,\mathbf{K})$ is defined in Eq.~(\ref{eq:I}) and is assumed to vary slowly near the point $\Omega=\Omega_v$ and to not have branch cuts. The expansion of $\varepsilon(\Omega,K)$ near $\Omega=\Omega_v$ is
\begin{equation}
\left.\varepsilon(\Omega,K)\right|_{\Omega\approx\Omega_v} = \varepsilon(\Omega_v,K) + \beta(K)(\Omega-\Omega_v) + O\left[(\Omega-\Omega_v)^2\right], \label{eq:epsilon_approx}
\end{equation}
with $\alpha(K_\parallel)$, $\beta(K_\parallel)$ and $\beta(K)$ defined in Eqs~(\ref{eq:alpha})--(\ref{eq:beta(K)}) of Appendix~\ref{app:rho_approximate} (note that in general $\beta(K)\neq\beta(K_\parallel)$, since $K\neq K_\parallel$). The approximation (\ref{eq:rho_approx}) with (\ref{eq:epsilon_approx}) is expressed in terms of elementary functions of $\Omega-\Omega_v$, which can be analytically continued into the lower vicinity ${\rm Im}(\Omega)<0$ of the branching point $\Omega=\Omega_v$. When doing so, the complex function $\sqrt{\Omega-\Omega_v}$ should be defined so that its branch cut goes vertically downwards from its branching point $\Omega=\Omega_v$. This branch cut can be parametrized as
\[
\Omega_{\rm cut}=+\Omega_v(K_\parallel) - i\eta,\ \ \eta\geq 0.
\]

The integral around the branch cut in the vicinity of the right branching point $\Omega=+\Omega_v$ is (the $(+)$ superscript denotes the right branch cut)
\begin{eqnarray}
\tilde\rho^{(+)}_\mathbf{K}(T) = \frac{i}{2\pi}e^{-i\Omega_vT}\int_0^\infty{d\eta e^{-\eta T}\left[\tilde\rho^{(+)}_L(\Omega_v-i\eta,\mathbf{K}) - \tilde\rho^{(+)}_R(\Omega_v-i\eta,\mathbf{K})\right]}, \label{eq:rho+_gen}
\end{eqnarray}
where $\tilde\rho^{(+)}_{L,R}(\Omega,\mathbf{K})$ are the left and right branches of the analytic continuation of (\ref{eq:rho_approx}) into the lower semiplane ${\rm Im}(\Omega)<0$.

%**** rewrite to accommodate the third case on p. 12 of my notes!
We first assume that the function $\varepsilon(\Omega,K)$, analytically continued to ${\rm Im}(\Omega)\leq0$, does not have branching points in some (perhaps small) vicinity of the point $\Omega=\Omega_v$; hence we have 
\[
\varepsilon_L(\Omega,K) = \varepsilon_R(\Omega,K)
\]
near $\Omega=\Omega_v$. Then the only function in $\tilde\rho^{(+)}_{L,R}(\Omega,\mathbf{K})$ with a cut is $\sqrt{\Omega-\Omega_v}$. Choosing its cut as specified above, we have
\begin{eqnarray}
\left(\sqrt{\Omega-\Omega_v}\right)_R = -\left(\sqrt{\Omega-\Omega_v}\right)_L = \sqrt{\left|\Omega-\Omega_v\right|}e^{-i\pi/4} = \sqrt{|\eta|}e^{-i\pi/4}.  \label{eq:sqrt}
\end{eqnarray}

Using (\ref{eq:rho_approx}), (\ref{eq:sqrt}) and (\ref{eq:epsilon_approx}), from (\ref{eq:rho+_gen}) we obtain
\begin{eqnarray}
\tilde\rho^{(+)}_\mathbf{K}(T) &=& \frac{ie^{i\pi/4}}{\pi}\sqrt{\alpha(K_\parallel)\beta(K_\parallel)}K_\parallel \left.I(\Omega_v,\mathbf{K})\right|_{K_x=0} e^{-i\Omega_v T} \nonumber \\
&\times& \int_0^\infty{d\eta \frac{e^{-\eta T} \sqrt{\eta}}{1 + i\eta\ \alpha(K_\parallel)\beta(K_\parallel)K_\parallel^2}\left[1-\frac{1}{\varepsilon(\Omega_v,K) - i\eta\beta(K) + O(\eta^2)}\right]}.  \label{eq:rho(T)}
\end{eqnarray}
The long-time asymptote of $\tilde\rho^{(+)}_\mathbf{K}(T)$ depends on whether $\varepsilon(\Omega_v,K)$ tends to zero or not, which in turn depends on the value of $K_x$. Below we consider the two cases: (i) $K_x\neq0$, so that $\varepsilon(\Omega_v,K)\neq0$, and (ii) $K_x\to0$, so that $\varepsilon(\Omega_v,K)\to 0$ [since $\Omega_v$ is the root of (\ref{eq:Omega_v})]. In these cases, (\ref{eq:rho(T)}) gives, respectively:
\begin{eqnarray}
\tilde\rho^{(+)}_{\mathbf{K}}(T)&\propto&\frac{e^{-i\Omega_vT}}{T^{3/2}} + O(T^{-5/2}),\ \ K_x\neq0, \label{eq:rho+_Kx!=0} \\
\tilde\rho^{(+)}_{\mathbf{K}}(T)&\propto&\frac{e^{-i\Omega_vT}}{T^{1/2}} + O(T^{-3/2}),\ \ K_x\to0.  \label{eq:rho+_Kx=0}
\end{eqnarray}
Similarly for the contribution of the left branch cut, $\Omega=-\Omega_v-i\eta,\ \eta\geq0$, we have $\tilde\rho^{(-)}_{\mathbf{K}}(T)$ given by Eqs~(\ref{eq:rho+_Kx!=0})--(\ref{eq:rho+_Kx=0}) with $\Omega_v$ replaced with $-\Omega_v$. The total contribution of both branch cuts $\tilde\rho^{\rm (cuts)}_{\mathbf{K}}(T)=\tilde\rho^{(+)}_{\mathbf{K}}(T)+\tilde\rho^{(-)}_{\mathbf{K}}(T)$ is then
\begin{eqnarray}
\tilde\rho^{\rm (cuts)}_{\mathbf{K}}(T)&\propto&\frac{\cos\left(\Omega_vT\right)}{T^{3/2}} + O(T^{-5/2}),\ \ K_x\neq0, \label{eq:rho(T)_cuts_Kx!=0} \\
\tilde\rho^{\rm (cuts)}_{\mathbf{K}}(T)&\propto&\frac{\cos\left(\Omega_vT\right)}{T^{1/2}} + O(T^{-3/2}),\ \ K_x\to0.  \label{eq:rho(T)_cuts_Kx=0}
\end{eqnarray}
Note that the frequency of these oscillations is equal to the frequency of volume plasma waves with the same wavelength, $K=K_\parallel$, and thus exceeds the frequency $\Omega_s$ of the surface oscillations (\ref{eq:rho(T)_poles}) due to the poles of $1/\zeta$.

Above we have assumed that the function $\varepsilon(\Omega,K)$, analytically continued to ${\rm Im}(\Omega)\leq0$, does not branch in at least some vicinity of the branching point $\Omega_v$ of $\zeta(\Omega,K_\parallel)$. However, this assumption is violated in a special case considered below. Indeed, the function $\varepsilon(\Omega,K)$ itself, when analytically continued into ${\rm Im}(\Omega)\leq0$, has branching points at $\Omega=\pm\sqrt{3}K\in\mathbb{R}$, with the branch cuts going downwards~\cite{Vlad_Tysh_UFN_2011}. Thus for $K_x\to\sqrt{\Omega_v^2(K_\parallel)/3 - K_\parallel^2}$ the branching points of $\varepsilon(\Omega,K)$ merge with the branching points of $\zeta(\Omega,K_\parallel)$, and their respective branch cuts merge at least in some lower vicinity of the coinciding branching points. In this case $\varepsilon_L(\Omega,K)\neq\varepsilon_R(\Omega,K)$, and the above calculation is modified; instead, we have for ${\rm Im}(\Omega)\leq0$
\begin{eqnarray}
\varepsilon_R(\Omega,K) &=& 1+\frac{1}{K^2}\left[1-\frac{\Omega}{2\sqrt{3}K}\log\left(\frac{\Omega+\sqrt{3}K}{\Omega-\sqrt{3}K}\right)\right], \\
\varepsilon_L(\Omega,K) &=& \varepsilon_R(\Omega,K) + \frac{i\pi\Omega}{\sqrt{3}K^3}
\end{eqnarray}
(we still assume that $I(\Omega,\mathbf{K})$ does not have branching points). Then, after some calculation, we obtain for $\tilde\rho^{(+)}_L(\Omega_v-i\eta,\mathbf{K}) - \tilde\rho^{(+)}_R(\Omega_v-i\eta,\mathbf{K})$ in (\ref{eq:rho+_gen}) in this case:
\begin{equation}
\tilde\rho^{(+)}_L(\Omega_v-i\eta,\mathbf{K}) - \tilde\rho^{(+)}_R(\Omega_v-i\eta,\mathbf{K}) = -i\left.\frac{I(\Omega_v,\mathbf{K})}{\varepsilon_R(\Omega_v,K)}\right|_{K_x=\sqrt{\Omega_v^2/3-K_\parallel^2}} + O\left(|\eta|^{1/2}\right).
\end{equation}
Carrying out the integration in (\ref{eq:rho+_gen}) and adding the similar contribution of the left cut, we finally obtain for the contribution of branch cuts in this special case:
\begin{equation}
\tilde\rho^{\rm (cuts)}_{\mathbf{K}}(T)\propto\frac{\cos(\Omega_v T)}{T} + O\left(T^{-3/2}\right),\ \ {\rm for\ } K_x\to\sqrt{\Omega_v^2/3-K_\parallel^2},  \label{eq:rho(T)_cuts_Kx=Kxc}
\end{equation}
which has the same frequency as (\ref{eq:rho(T)_cuts_Kx!=0})--(\ref{eq:rho(T)_cuts_Kx=0}), but a different temporal attenuation exponent.

\section{Discussion}
We thus see that our system supports two distinct types of surface oscillations, with different frequencies and temporal attenuation: (i) exponentially damped surface oscillations (\ref{eq:rho(T)_poles}) with frequency $\Omega_s(K_\parallel)$, due to the poles of $\tilde\rho(\Omega,\mathbf{K})$~\cite{Tyshetskiy_etal_PoP_2012}; and (ii) power-law attenuated surface oscillations (\ref{eq:rho(T)_cuts_Kx!=0}), (\ref{eq:rho(T)_cuts_Kx=0}), and (\ref{eq:rho(T)_cuts_Kx=Kxc}) with frequency $\Omega_v(K_\parallel)>\Omega_s(K_\parallel)$, due to the branch cuts of $\tilde\rho(\Omega,\mathbf{K})$. Since the power-law attenuation is slower than the exponential attenuation, these oscillations should become dominant at large times, and should become observable in principle, e.g., by analyzing a spectrum of the reflected light in experimental setups for excitation of surface waves in thin metal films by an incident light using Otto or Kretschmann configurations~\cite{Otto_1968,Kretschmann_1968}. At small values of $K_z$, the frequency difference between these two types of surface oscillations approaches a third of the metal's plasma frequency, $\Omega_v(K_\parallel)-\Omega_s(K_\parallel)\to 1-1/\sqrt{2}\approx 0.3$, and thus the absorption lines in the reflected light spectrum, corresponding to excitation of surface waves of these two types, should be clearly separated and detectable.

It is interesting to note that, as seen from (\ref{eq:rho(T)_cuts_Kx!=0}), (\ref{eq:rho(T)_cuts_Kx=0}), and (\ref{eq:rho(T)_cuts_Kx=Kxc}), different $K_x$ components in the wave packet, making up the field of the surface oscillation of this type, are attenuated at different rates. Since $T^{-1/2}$ decays slower than $T^{-1}$ or $T^{-3/2}$, the small-$K_x$ part of the wave packet becomes dominant over the large-$K_x$ part at large times. This corresponds to penetration of the charge perturbation away from the surface and deeper into plasma.

We should stress that the reported prediction of the new type of surface waves due to the contribution of cuts of $\zeta(\Omega,K_\parallel)$ is a result of rigorous solution of the initial value problem, and would not have been possible to make by just seeking for solutions of the Vlasov equation (\ref{eq:Vlasov}) in the form $f_p\propto \exp\left(i \mathbf{K_\parallel\cdot R_\parallel} - i\Omega T \right)$ (in fact, the latter would be conceptually wrong, as discussed in Ref.~\cite{Komp_etal_2013}).

The presented analysis relies on several assumptions discussed in detail in Ref.~\cite{Tyshetskiy_etal_PoP_2012}, of which perhaps the most critical ones are the neglect of the quantum recoil (which does not play a significant role unless the wavelengths are extremely short), and the assumptions of collisionless plasma and of the sharp perfectly reflecting boundary confining the plasma. Relaxing the first two assumption should not change the results qualitatively~\cite{Tyshetskiy_etal_PoP_2012(2)}, and there should still be two types of the surface waves when the quantum recoil is retained in the model. Indeed, with quantum recoil retained, the function $1/\zeta(\Omega,K_\parallel)$ will still have the branch cuts in the lower semi-plane of complex $\Omega$, whose contribution would still lead to the second type of surface oscillations reported here. Yet relaxing the assumption of the sharp plasma boundary may affect the results obtained here in a non-trivial way. Firstly, the smooth boundary leads to a new resonant damping of surface oscillations, significantly increasing the exponential damping rate $|\Gamma_s|$ in (\ref{eq:rho(T)_poles})~\cite{Marklund_etal_new_quantum_limits}. Secondly, allowing for boundary smoothness (with a simultaneous account for the quantum tunneling, as they both have the same spatial scales) should change the analytic properties of $\tilde\rho(\Omega,\mathbf{K})$ in the lower semiplane ${\rm Im}(\Omega)<0$ of complex $\Omega$ plane, and thus may change its branch cuts and their contribution into (\ref{eq:inv_Laplace}). The generalization of this study to the case of non-sharp plasma boundary is however beyond the scope of this paper and is left for future work.

\section{Summary}
We have studied the temporal evolution of initial perturbation of a semi-bounded degenerate quantum plasma with a sharp boundary, in the electrostatic limit. By rigorously solving the initial value problem for the set of coupled Vlasov-Poisson equations describing the system kinetics, we have found that the part of electric potential corresponding to surface waves can be represented, at large times, as a sum of two terms, one corresponding to ``conventional'' surface wave, the other corresponding to a new type of surface waves. This new surface wave has a larger frequency than the ``conventional'' surface wave (in fact, its frequency corresponds to the frequency of a volume plasmon with the same wavelength, $K=K_\parallel$), and a slower temporal attenuation (power-law attenuation versus the exponential damping of the ``conventional'' wave), making it dominant at large times. These two types of surface waves should in principle be detectable as separate waves in sensitive enough experiments on exciting surface waves in a metal film by an incident light beam (using Otto or Kretschmann configurations). The new type of surface waves predicted here may prove to be important for designing future plasmonic devices and technologies employing interaction of light with collective surface modes. 

\acknowledgments{This work was supported by the Australian Research Council.}

\appendix
\section{Branching points of $\zeta(\Omega,K_\parallel)$ \label{app:branching}}
Let us show that the function $\zeta(\Omega,K_\parallel)$, analytically continued into ${\rm Im}(\Omega)\leq0$, has branching points at $\Omega=\pm\Omega_v(K_\parallel)\in\mathbb{R}$, with the corresponding branch cuts going down from these two points. We start from $\zeta(\Omega,K_\parallel)$ defined by Eq.~(\ref{eq:zeta}) for ${\rm Im}(\Omega)>0$, and then continuously change ${\rm Im}(\Omega)$ to negative values. In this process, we must consider how the singularities of the function $\left[K^2\varepsilon(\Omega,K)\right]^{-1}$ under the integral in (\ref{eq:zeta}) change in the complex $K_x$ plane~\cite{Tyshetskiy_etal_PoP_2012}. These singularities are:
\begin{enumerate}
\item Branch cuts of the complex square root $\sqrt{K_x^2+K_\parallel^2}$, defined by two parametric equations:
\begin{equation}
K_x = \pm i \sqrt{K_\parallel^2 + \tau},\ {\rm with\ }K_\parallel>0,\ \tau\in[0,+\infty). \label{eq:sqrt_cuts}
\end{equation}
\item Branch cut of the complex logarithm in (\ref{eq:epsilon_degenerate}), taken along the negative real axis of the argument $(\Omega+\sqrt{3}K)/(\Omega-\sqrt{3}K)$. This branch cut maps into two branch cuts of $\left[K^2\ \varepsilon(\Omega,K)\right]^{-1}$ in the complex $K_x$ plane, given by two parametric equations:
\begin{equation}
K_x = \pm i \sqrt{\frac{\Omega^2}{3}\left(\frac{\tau+1}{\tau-1}\right)^2 - K_\parallel^2},\ {\rm with\ }\Omega\in\mathbb{C},\ K_\parallel>0,\ \tau\in[0,+\infty).  \label{eq:log_cuts}
\end{equation}
\item Two poles $K_x=\pm i K_\parallel$ ($K_\parallel>0$) at the roots of $K^2=0$, lying symmetrically above and below the real axis of the complex $K_x$ plane.
\item Two poles $\pm K_x^{r}\in\mathbb{C}$ at the roots of $\varepsilon(\Omega,K)=0$. Note that for any $K\in\mathbb{R}$, $\varepsilon(\Omega,K)=0$ does not have roots with ${\rm Im}(\Omega)>0$, if the plasma equilibrium is stable~\cite{Penrose_1960}, which is the case considered here; therefore, for any ${\rm Im}(\Omega)>0$ the poles $\pm K_x^{r}$ are located \textit{away} from the real axis of the complex $K_x$ plane, and thus do not lie on the integration contour in (\ref{eq:zeta}). 
\end{enumerate}
\begin{figure}
\centerline{\includegraphics[width=2.2in]{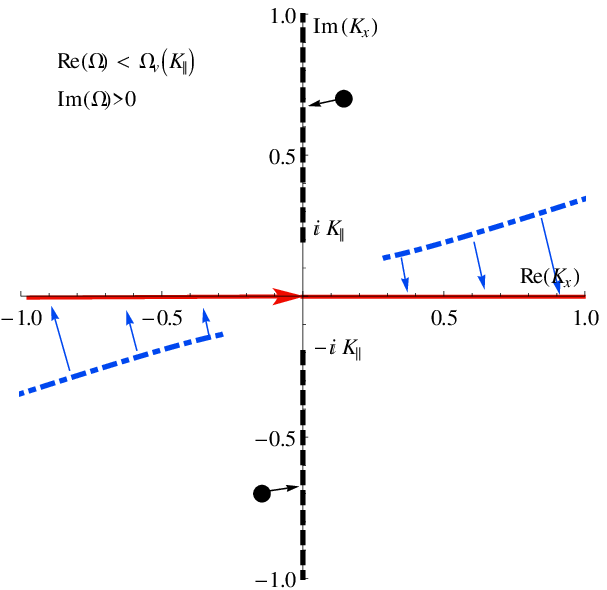}\ \ \ \includegraphics[width=2.2in]{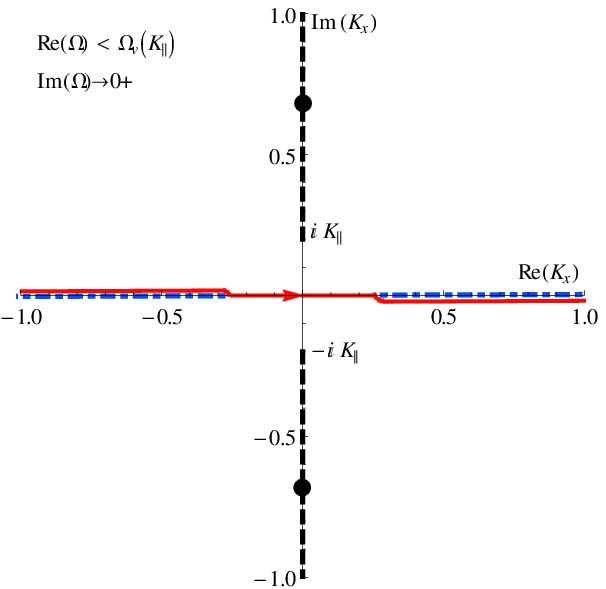}} 
\centerline{\includegraphics[width=2.2in]{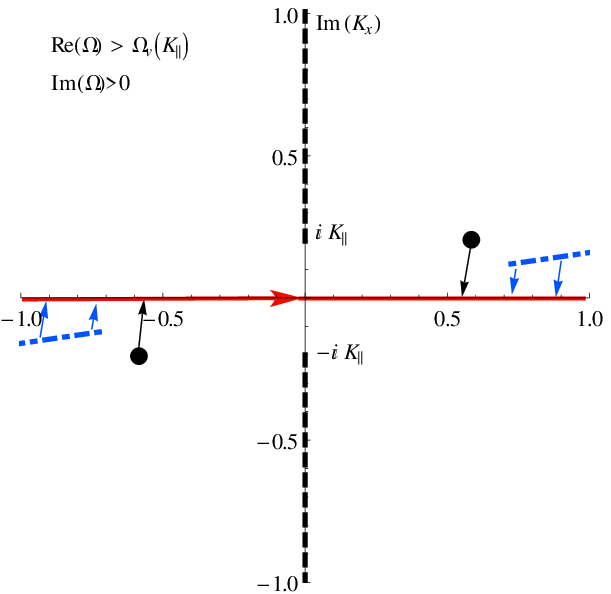}\ \ \ \includegraphics[width=2.2in]{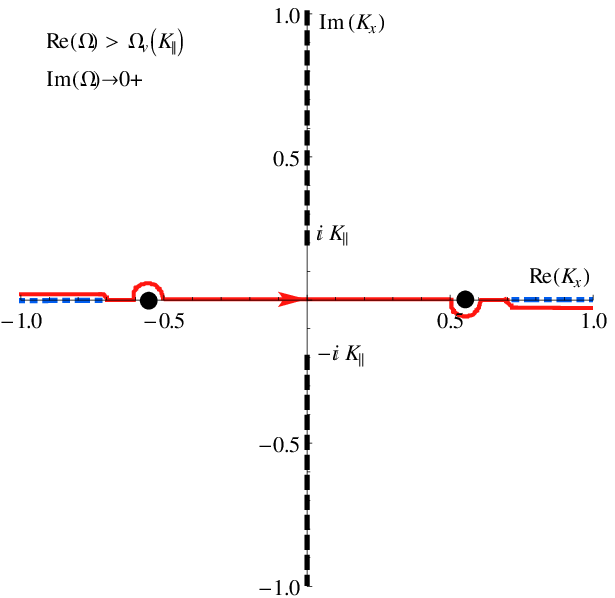}}
\caption{\label{fig:Kx_plane} (Color online) Singularities of the function $\left[K^2\ \varepsilon(\Omega,K)\right]^{-1}$ in the complex $K_x$ plane, and their modification from ${\rm Im}(\Omega)>0$ (left panels) to the limit ${\rm Im}(\Omega)\rightarrow0+$ (right panels), for $0<{\rm Re}(\Omega)<\Omega_v(K_\parallel)$ (upper row) and ${\rm Re}(\Omega)>\Omega_v(K_\parallel)>0$ (lower row), with $\Omega_v(K_\parallel)$ defined from (\ref{eq:Omega_v}). The branch cuts (\ref{eq:sqrt_cuts}) and (\ref{eq:log_cuts}) are shown with the black dashed lines and the blue dot-dashed lines, respectively. The poles $\pm K_x^{r}$ where $\varepsilon(\Omega,K)=0$ are shown with the filled circles. The arrows show the direction of motion of the singularities when ${\rm Im}(\Omega)\rightarrow0+$. The contour of $K_x$ integration in (\ref{eq:zeta}) is shown with the solid red line.}
\end{figure}

In the process of analytic continuation, as ${\rm Im}(\Omega)\to 0+$, these singularities deform/move in the complex $K_x$ plane, as shown in Fig.~\ref{fig:Kx_plane}. We have the following cases: (i) $|{\rm Re}(\Omega)|<|\Omega_v(K_\parallel)|$, and (ii) $|{\rm Re}(\Omega)|>|\Omega_v(K_\parallel)|$, where $\pm\Omega_v(K_\parallel)\in\mathbb{R}$ are defined by the equation~(\ref{eq:Omega_v}). The difference between these two cases is that for $|{\rm Re}(\Omega)|>|\Omega_v(K_\parallel)|$, the poles $\pm K_x^{r}$ cross the real axis in $K_x$ plane and deform the integration contour when ${\rm Im}(\Omega)\to 0+$ and beyond to negative values, while for $|{\rm Re}(\Omega)|<|\Omega_v(K_\parallel)|$ they do not. Thus we have that in these two cases, the integration contours in the function $\zeta(\Omega,K_\parallel)$ continued to ${\rm Im}(\Omega)<0$ are \textit{different}, and thus the values of $\zeta(\Omega,K_\parallel)$ in the lower semiplane of complex $\Omega$ are also different for $|{\rm Re}(\Omega)|<|\Omega_v(K_\parallel)|$ and for $|{\rm Re}(\Omega)|>|\Omega_v(K_\parallel)|$. Hence, the points $\Omega=\pm\Omega_v(K_\parallel)\in\mathbb{R}$ separating these two cases must necessarily be the branching points of $\zeta(\Omega,K_\parallel)$ in ${\rm Im}(\Omega)\leq0$, with branch cuts (separating the different values of the analytically continued $\zeta$) going down into the lower semiplane of complex $\Omega$. At the branching points $\pm\Omega_v(K_\parallel)$, the function $\zeta(\Omega,K_\parallel)$ has a singularity~\cite{Tyshetskiy_etal_PoP_2012}.

\section{Approximation of $\tilde\rho(\Omega,\mathbf{K})$ in the upper vicinity of the branching point $\Omega=+\Omega_v(K_\parallel)$ \label{app:rho_approximate}}
The function $\tilde\rho(\Omega,\mathbf{K})$ defined in (\ref{eq:rho(w,k)}) contains integrals of the form
\begin{equation}
\int_{-\infty}^{+\infty}\frac{dK_x}{K^2 \varepsilon(\Omega,K)}A(\Omega,K) = \frac{1}{K_\parallel^2}\int_{-\infty}^{+\infty}\frac{A(\Omega,K) dK_x}{\left(1+K_x^2/K_\parallel^2\right) \varepsilon(\Omega,K)},  \label{eq:int_Kx}
\end{equation}
which need to be approximated in the upper vicinity of the branching point $\Omega=\Omega_v$. The main contribution into the integrals (\ref{eq:int_Kx}) is from the vicinity of $K_x=0$. The expansion of $\varepsilon(\Omega,K)$ near $\Omega=\Omega_v$, $K_x=0$ under the integral is
\begin{eqnarray}
\varepsilon(\Omega,K)=\left(\Omega-\Omega_v\right)\beta(K_\parallel) + \alpha(K_\parallel)K_x^2 + O\left[\left(\Omega-\Omega_v\right)^2\right] + O\left(K_x^4\right),
\end{eqnarray}
where
\begin{eqnarray}
\alpha(K_\parallel) &=& \frac{1}{2}\left.\varepsilon^{\prime\prime}_{Kx,Kx}(\Omega,K)\right|_{K_x=0,\ \Omega=\Omega_v}, \label{eq:alpha}\\
\beta(K_\parallel) &=& \left.\varepsilon^\prime_\Omega(\Omega,K)\right|_{K_x=0,\ \Omega=\Omega_v}, \\
\beta(K) &=& \left.\varepsilon^\prime_\Omega(\Omega,K)\right|_{\Omega=\Omega_v},  \label{eq:beta(K)}
\end{eqnarray} 
and the primes denoting partial derivatives with respect to the corresponding variables, e.g., $\varepsilon^\prime_\Omega=\partial\varepsilon/\partial\Omega$. Here we have taken into account that $\varepsilon(\Omega_v,K_\parallel)=0$ (by definition of $\Omega_v$) and $\left.\varepsilon^\prime_{K_x}(\Omega,K)\right|_{K_x=0} = \left.\left[\left(K_x/K\right)\varepsilon^\prime_K\right]\right|_{K_x=0} = 0$.

The function $\zeta(\Omega,K_\parallel)$ in the upper vicinity of $\Omega=\Omega_v(K_\parallel)$ is then
\begin{eqnarray}
\zeta(\Omega,K_\parallel) &\approx& \frac{1}{2}+\frac{1}{2\pi K_\parallel}\int_{-\infty}^{+\infty}{\frac{dK_x}{\left(1+K_x^2/K_\parallel^2\right)\left[\beta(K_\parallel)(\Omega-\Omega_v) + \alpha(K_\parallel) K_x^2 \right]}} \nonumber \\
&=& \frac{1}{2} + \frac{1}{2K_\parallel\sqrt{\alpha(K_\parallel)\beta(K_\parallel)}}\frac{1}{\sqrt{\Omega-\Omega_v}}. \label{eq:zeta_approx}
\end{eqnarray}
Here we neglected $O[(\Omega-\Omega_v)^2]$ (as we are considering a small vicinity of $\Omega=\Omega_v$), $O(K_x^4)$ (as the main contribution into the integral is from $K_x\approx0$), and $O[(\Omega-\Omega_v)K_x^2]$ (due to the combination of the above two reasons). 
Similarly, in the upper vicinity of $\Omega=\Omega_v(K_\parallel)$ we obtain
\begin{eqnarray}
\int_{-\infty}^{+\infty}{dK_x^\prime\frac{I(\Omega,\mathbf{K^\prime})}{{K^\prime}^2\varepsilon(\Omega,K^\prime)}} \approx \frac{\pi \left.I(\Omega_v,\mathbf{K})\right|_{K_x=0}}{\sqrt{\alpha(K_\parallel)\beta(K_\parallel)} K_\parallel^2 \sqrt{\Omega-\Omega_v}},   \label{eq:I_approx}
\end{eqnarray}
where we have assumed that the function $I(\Omega,\mathbf{K})$ varies slowly near the point $\Omega=\Omega_v$, and does not have branch cuts.

Combining (\ref{eq:zeta_approx}) and (\ref{eq:I_approx}) in (\ref{eq:rho(w,k)}), we obtain the approximation (\ref{eq:rho_approx}) for $\tilde\rho(\Omega,\mathbf{K})$ in the upper vicinity of $\Omega=\Omega_v$.

%%%\bibliography{paper_cuts}

\end{document}